\documentclass[aps,prl,twocolumn,superscriptaddress]{revtex4-2}
\setcitestyle{sort,compress=true}
\usepackage[T1]{fontenc}
\usepackage{xcolor}
\usepackage{graphicx}
\usepackage{dcolumn}
\usepackage{bm}
\usepackage{amsmath}
\usepackage[colorlinks = true,]{hyperref}
\usepackage{siunitx}
\usepackage{tikz}

\begin{document}

\title{Collective Strong Coupling of Thermal Atoms to Integrated Microring Resonators}

\author{Xiaoyu~Cheng}
\email{xcheng@pi5.physik.uni-stuttgart.de}
\affiliation{5th Institute of Physics and Center for Integrated Quantum Science and Technology, University of Stuttgart, Pfaffenwaldring 57, 70569 Stuttgart, Germany}
\author{Benyamin~Shnirman}
\affiliation{5th Institute of Physics and Center for Integrated Quantum Science and Technology, University of Stuttgart, Pfaffenwaldring 57, 70569 Stuttgart, Germany}
\affiliation{Institut für Mikroelektronik Stuttgart (IMS CHIPS), Allmandring 30a, 70569 Stuttgart, Germany}

\author{Alexandra~Köpf}
\affiliation{5th Institute of Physics and Center for Integrated Quantum Science and Technology, University of Stuttgart, Pfaffenwaldring 57, 70569 Stuttgart, Germany}
\affiliation{Institut für Mikroelektronik Stuttgart (IMS CHIPS), Allmandring 30a, 70569 Stuttgart, Germany}

\author{Guangcanlan~Yang}
\affiliation{Department of Electrical and Computer Engineering, Yale University, New Haven, CT 06511, USA}


\author{Hong~X.~Tang}
\affiliation{Department of Electrical and Computer Engineering, Yale University, New Haven, CT 06511, USA}

\author{Hadiseh~Alaeian}
\affiliation{Elmore Family School of Electrical and Computer Engineering, Purdue University, West Lafayette, Indiana 47907, USA}
\affiliation{Department of Physics and Astronomy, Purdue University, West Lafayette, Indiana 47907, USA}

\author{Tilman~Pfau}
\affiliation{5th Institute of Physics and Center for Integrated Quantum Science and Technology, University of Stuttgart, Pfaffenwaldring 57, 70569 Stuttgart, Germany}
\author{Robert~Löw}
\affiliation{5th Institute of Physics and Center for Integrated Quantum Science and Technology, University of Stuttgart, Pfaffenwaldring 57, 70569 Stuttgart, Germany}

\date{\today}

\begin{abstract}
  Strong coupling between atomic ensembles and high-quality optical cavities enables collective and nonlinear phenomena that are central to cavity quantum electrodynamics (cQED). Although many experiments have been performed on this topic, most of them have focused on cold atoms. Here, we experimentally demonstrate collective strong coupling between thermal rubidium (Rb) vapor and high-quality silicon nitride microring resonators (MRRs) on an integrated photonic chip. We observe cavity mode splitting, with a measured collective coupling strength of $g_N/2\pi\approx\SI{1}{\giga\hertz}$ and a collective cooperativity of $C_N\approx2$ at $\SI{110}{\celsius}$, indicating coherent energy exchange between the atomic ensemble and the cavity mode despite rapid decoherence in the thermal vapor system. We infer an average of $20$ atoms participating in the collective interaction, yielding a single-atom cooperativity of $C_0=0.1$ and approaching the single-atom strong-coupling regime. Our results establish the integrated thermal vapor MRR platform as a robust, compact, and scalable system for studying collective and nonlinear phenomena in cQED.
\end{abstract}

\maketitle

Research in cQED dates back to the 1940s, when Purcell showed that an emitter's radiative properties can be modified by its environment~\cite{Purcell}, particularly by an optical cavity. Theoretical descriptions of single-atom and many-atom cavity coupling followed in the 1960s~\cite{Jaynes,Tavis}, and strong-coupling signatures such as mode splitting were later observed with atomic beams~\cite{CarmichaelNMSplitting1,CarmichaelNMSplitting2,RempeNMSplitting}, cold atoms, and trapped ions~\cite{HAROCHE1985347,PhysRevLett.93.233603,PhysRevLett.94.033002,PhysRevLett.124.013602,doi:10.1126/science.287.5457.1447,doi:10.1126/science.1095232,PhysRevLett.130.173601}.
Recent advances in microfabrication have enabled high-finesse microcavities with small mode volumes, further enhancing atom--cavity interactions, with many strong-coupling demonstrations across different platforms such as miniaturized Fabry-P\'erot cavities~\cite{Colombe2007Nature_StrongCouplingBEC}, micro photonic-crystal cavities~\cite{PhysRevLett.124.063602}, microbottle resonators~\cite{PhysRevLett.126.233602,PhysRevLett.110.213604}, microtoroidal resonators~\cite{Aoki_2006} and microring resonators~\cite{ 10.1063/5.0023464}.

Despite these major successes of strong-coupling cQED with cold atoms and nanophotonic cavities, an important question has remained open: how to combine strong light--matter interaction with scalability, compactness, robustness, and manufacturability. Traditional cold-atom platforms often offer outstanding performance, but they are also mechanically delicate, experimentally demanding, bulky, and difficult to scale to large numbers of devices. These limitations have motivated growing interest in more integrated platforms that can be implemented in a more compact, robust, and scalable format. One particularly promising direction is to combine nanophotonic microcavities with thermal atomic vapors, which offer a much simpler and more compact alternative to laser-cooled atomic systems.
Recent work has therefore explored the feasibility of achieving strong coupling between thermal atoms and microcavities~\cite{Alaeian2020, Stern:12, Ritter_2016}. Despite the transient decoherence and Doppler broadening associated with thermal atoms, experiments have nevertheless demonstrated collective strong coupling between thermal atomic vapor and high-quality microcavities~\cite{PhysRevA.57.R2293, Zektzer:24,Naiman2021ResearchSquare_LargeCooperativity}. In addition to reduced experimental complexity, the compact footprint and scalability of such platforms make them promising candidates for extension from single-cavity devices to coupled multi-cavity systems, potentially enabling on-chip quantum networks with rapid and localized reconfigurability.

In this letter, we demonstrate collective strong coupling between thermal rubidium vapor and integrated MRRs, evidenced by a mode splitting of $\SI{2}{\giga\hertz}$, corresponding to collective coupling strength $g_N/2\pi\approx\SI{1}{\giga\hertz}$ and a collective cooperativity of $C_N\approx2$. We estimate that $\approx20$ Rb atoms contribute to the collective coupling, corresponding to a single-atom cooperativity of $C_0\approx0.1$.  By measuring the splitting systematically at different Rb vapor densities, we find good agreement with the theoretical prediction for collective coupling strength $g_N=\sqrt{N}g_0$~\cite{Tavis}, where $N$ is the average number of Rb atoms involved in the collective interaction and $g_0$ is the single-atom coupling strength. We also theoretically simulate the system with a simplified Tavis-Cummings model within the input–output framework that considers Doppler broadening and transient effects. The result of the simulation agrees well with the measurement.

\begin{figure}[htbp]
  \centering
  \includegraphics[width=0.95\columnwidth]{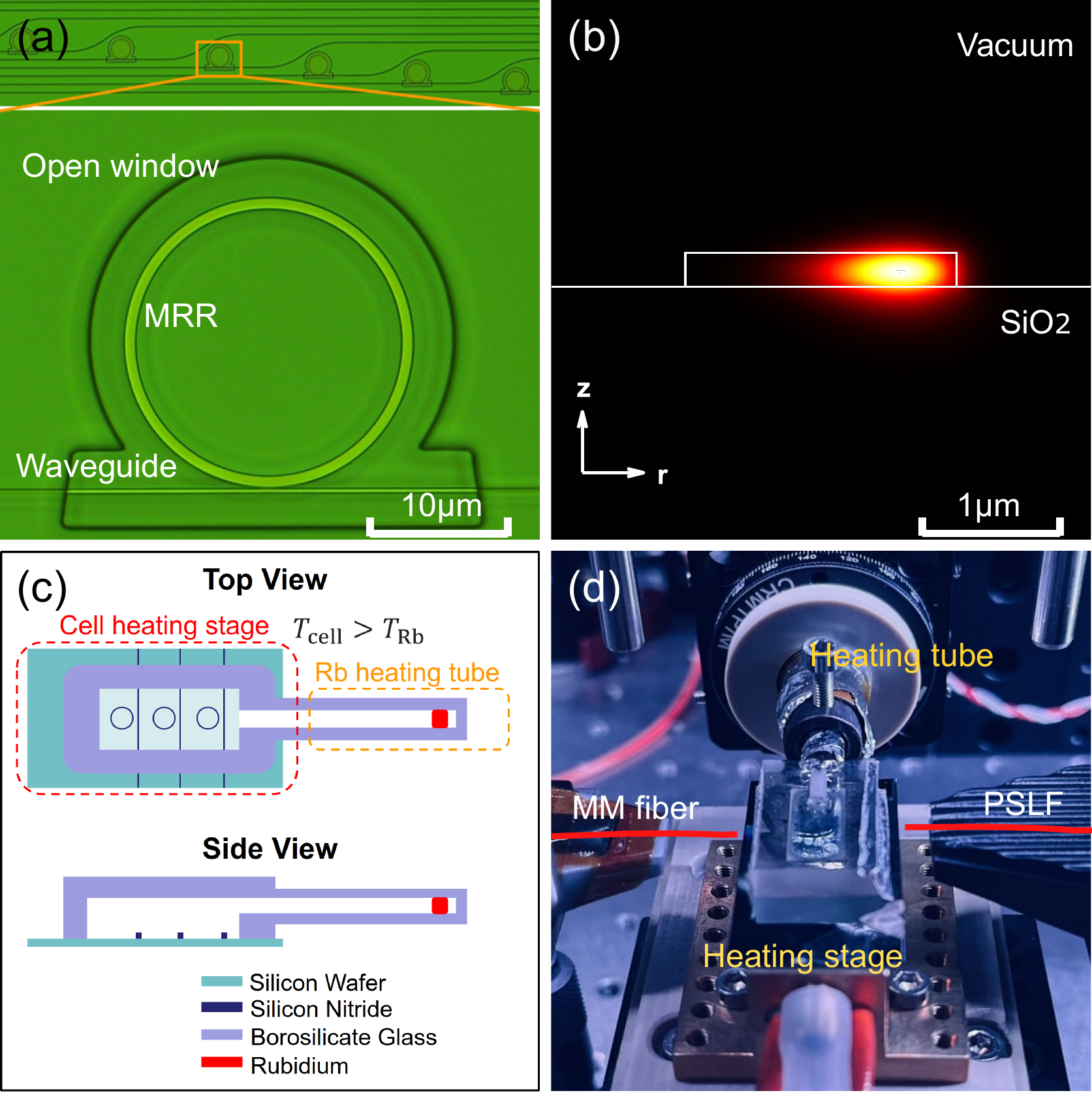}
  \caption{
    \textbf{(a)} Optical microscope image of a MRR group. The zoomed-in view shows a MRR with radius of \SI{10}{\micro\meter} coupled to a bus waveguide. The dark edge marks the opened window (image provided by IMS CHIPS).
    \textbf{(b)} COMSOL-simulated cross-sectional mode-field distribution of the TE$_{00}$ mode of the MRR in (a).
    \textbf{(c)} Schematic of the anodic bonded vapor cell: Rb is stored and heated in the glass reservoir tube, which is maintained at a lower temperature than the cell.
    \textbf{(d)} Image of the vapor cell mounted in the heating systems. A polarization-maintaining single-mode lensed fiber (PSLF) couples light into the waveguide and a multimode (MM) fiber collects the transmission light, both fibers highlighted in red.
  }
\label{fig:chipscells}
\end{figure}


\emph{Experimental setup.—} As a key part of the experiment, we design and fabricate ring resonators from a $\SI{250}{\nano\meter}$-thick Si$_3$N$_4$ layer deposited on thermally grown SiO$_2$ substrate on a Si wafer by low-pressure chemical vapor deposition (LPCVD). The ring radius is varied from \SI{13}{\micro\meter} to \SI{24}{\micro\meter}. The ring width is set to \SI{2}{\micro\meter} to confine the mode in the radial direction to reduce the scattering loss from sidewalls of the ring. A \SI{1}{\micro\meter}-thick thermal SiO$_2$ cladding layer is then deposited on top of the structured Si$_3$N$_4$ layer. Finally, the cladding above the ring resonator and the waveguide coupling region is locally removed (``opened'', Fig.~\ref{fig:chipscells}(a)) to enable interaction between Rb atoms and the MRRs. Figure~\ref{fig:chipscells}(b) shows the simulated cross-sectional electric-field distribution of the fundamental TE$_{00}$ mode in the MRR. From the mode distribution we extract the effective refractive index $n_{\mathrm{eff}}=1.81$, the evanescent-field decay length $\delta_{1/e}\approx\SI{90}{\nano\meter}$, and the mode volume $V_{\mathrm{mode}}=\SI{10.1}{\micro\meter^3}$ (see \hyperref[AppendixB]{Appendix~B}).
To load Rb vapor into the vicinity of the MRR, we anodically bond the photonic chip to a borosilicate glass cell. A high-purity metallic rubidium source with natural isotopic composition ($\approx\SI{72}{\percent}\,~^{85}\mathrm{Rb}$, $\approx\SI{28}{\percent}\,~^{87}\mathrm{Rb}$) is then transferred and hermetically sealed into the reservoir tube of the bonded cell (Fig.~\ref{fig:chipscells}(c)) under vacuum conditions (pressure below $\SI{1e-7}{\milli\bar}$). To prevent Rb condensation on the device, the vapor cell is heated by two independent heating systems, with the cell temperature maintained at least $\SI{20}{\celsius}$ above the Rb reservoir temperature.
Figure~\ref{fig:chipscells}(d) shows the separate heating systems together with the fiber-coupling setup of the photonic chip. Details of the measurement scheme are provided in \hyperref[AppendixA]{Appendix~A}, Fig.~\ref{fig:setup}.
\begin{figure}[t]
  \centering
  \includegraphics[width=0.75\columnwidth]{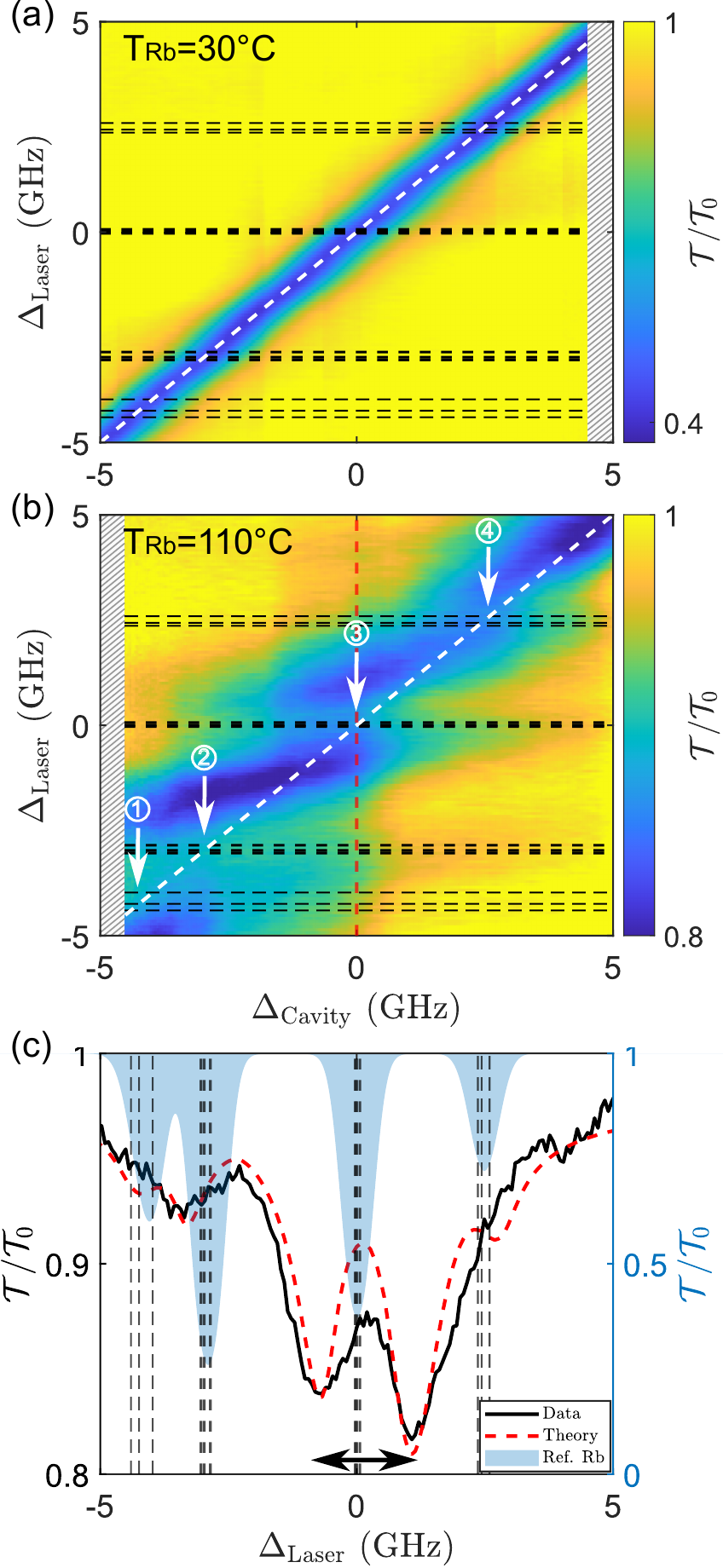}
  \caption{
    \textbf{(a), (b)} Two-dimensional colormaps of the transmission spectrum at $T_{\mathrm{Rb}}=\SI{30}{\celsius}$ and $\SI{110}{\celsius}$. Colorbars show the normalized transmission $\mathcal{T}/\mathcal{T}_0$. The cavity-detuning axes are aligned; hatched regions indicate missing data. Horizontal black dashed lines denote the hyperfine transitions of $^{85}\mathrm{Rb}$ and $^{87}\mathrm{Rb}$; the white dashed line marks the MRR resonance. In Fig.(b), four white arrows mark the center-of-mass positions of the expected avoided crossings, and the red dashed line indicates the selected transmission slice near the $^{85}\mathrm{Rb}\ 5^2S_{1/2}\ F_g=2 \to 5^2P_{3/2}\ F_e=2$ transition, presented in (c). In Fig.(c) the red dashed curve shows the theoretical prediction; the blue shaded region indicates the reference Rb vapor cell spectrum at the same $T_{\mathrm{Rb}}$; black dashed lines denote the same hyperfine transitions as in (a) and (b) and black double-arrow indicates the mode splitting.}
\label{fig:2Dcolormap&slicing}
\end{figure}


\emph{Measurement and Results.—}
We repeatedly scan the probe laser every $\SI{10}{\milli\second}$ while the MRR resonance is slowly swept across the Rb $\mathrm{D_2}$ transitions at a rate of $\SI{0.4}{\giga\hertz/\second}$. To achieve this, a tightly focused IR laser is directed onto the MRR and intensity modulated, providing localized tuning of the cavity resonance. Continuous acquisition of the probe transmission during this sweep yields the two-dimensional transmission map shown in Fig.~\ref{fig:2Dcolormap&slicing}.
At $T_{\mathrm{Rb}} = \SI{30}{\celsius}$, the Rb vapor density near the MRR is negligible (Fig.~\ref{fig:2Dcolormap&slicing}(a)). From the bare-cavity spectrum, we extract a loaded quality factor $Q_{\mathrm{load}} = \omega_{\mathrm{c}}/\kappa = 3 \times 10^5$ and finesse $\mathcal{F} = 1700$, with $\kappa/2\pi = \SI{1.2}{\giga\hertz}$ the FWHM. We further infer an intrinsic loss rate $\kappa_{\mathrm{i}}/2\pi = \SI{1}{\giga\hertz}$ from $Q_{\mathrm{int}} = \omega_{\mathrm{c}}/\kappa_{\mathrm{i}} = 2Q_{\mathrm{load}}/(1+\sqrt{\mathcal{T}_0})$, $\mathcal{T}_0$ being the resonance transmission of the MRR, and an external coupling rate $\kappa_{\mathrm{e}}/2\pi = \SI{230}{\mega\hertz}$~\cite{Liu:18}.

At higher Rb density, avoided crossings emerge for $T_{\mathrm{Rb}} \gtrsim \SI{80}{\celsius}$. The result at $T_{\mathrm{Rb}} = \SI{110}{\celsius}$ is shown in Fig.~\ref{fig:2Dcolormap&slicing}(b), in which we expect four avoided crossings at the position indicated by numbered white arrows. Based on the hyperfine structure of $^{85}\mathrm{Rb}$ and $^{87}\mathrm{Rb}$ (Fig.~\ref{fig:Rb_levelScheme}), the excited-state splittings of the $5^2P_{3/2}$ manifold are smaller than the Doppler ($\sim \SI{580}{\mega\hertz}$) and transit-time ($\sim \SI{280}{\mega\hertz}$) broadenings at this temperature. As a result, transitions from the same ground hyperfine state merge into a single broadened avoided crossing, yielding one feature per ground state.
However, only three distinct avoided crossings are observed. This is because the ground states $^{85}\mathrm{Rb}\ 5^2S_{1/2}\ F=3$ (arrow \tikz[baseline=(char.base)]{\node[draw, circle, inner sep=1pt] (char) {1};}) and $^{87}\mathrm{Rb}\ 5^2S_{1/2}\ F=2$ (arrow \tikz[baseline=(char.base)]{\node[draw, circle, inner sep=1pt] (char) {2};}) are separated by only $\sim \SI{1.2}{\giga\hertz}$, much smaller than the $>\SI{2}{\giga\hertz}$ separation to the other ground states, causing their corresponding avoided crossings to merge.

We obtain the mode-splitting spectrum at $T_{\mathrm{Rb}} = \SI{110}{\celsius}$ by slicing Fig.~\ref{fig:2Dcolormap&slicing}(b) along the red dashed line at arrow \tikz[baseline=(char.base)]{\node[draw, circle, inner sep=1pt] (char) {3};}, defined as zero cavity detuning. The measured splitting is $\approx \SI{2}{\giga\hertz}$, corresponding to a collective coupling strength of $g_N/2\pi \approx \SI{1}{\giga\hertz}$. From the evanescent mode volume and vapor density, we estimate $N \approx 20$ participating $^{85}\mathrm{Rb}$ atoms at $\SI{110}{\celsius}$ (see \hyperref[AppendixB]{Appendix~B}).
The splitted spectrum is asymmetric about the atomic transition due to off-resonant collective coupling to the $^{85}\mathrm{Rb}\ 5^2S_{1/2}\ F_g=3 \to 5^2P_{3/2}\ F_e=2,3,4$ transitions. This is supported by our theoretical simulations including all hyperfine transitions of the Rb $\mathrm{D_2}$ manifold (Fig.~\ref{fig:Theoretical_Simulation} in \hyperref[AppendixC]{Appendix~C}). The simulated transmission spectrum at zero cavity detuning at $T_{\mathrm{Rb}} = \SI{110}{\celsius}$ is also shown as the red dashed curve in Fig.~\ref{fig:2Dcolormap&slicing}(c).

\begin{figure}[htbp]
  \centering
  \includegraphics[width=0.75\columnwidth]{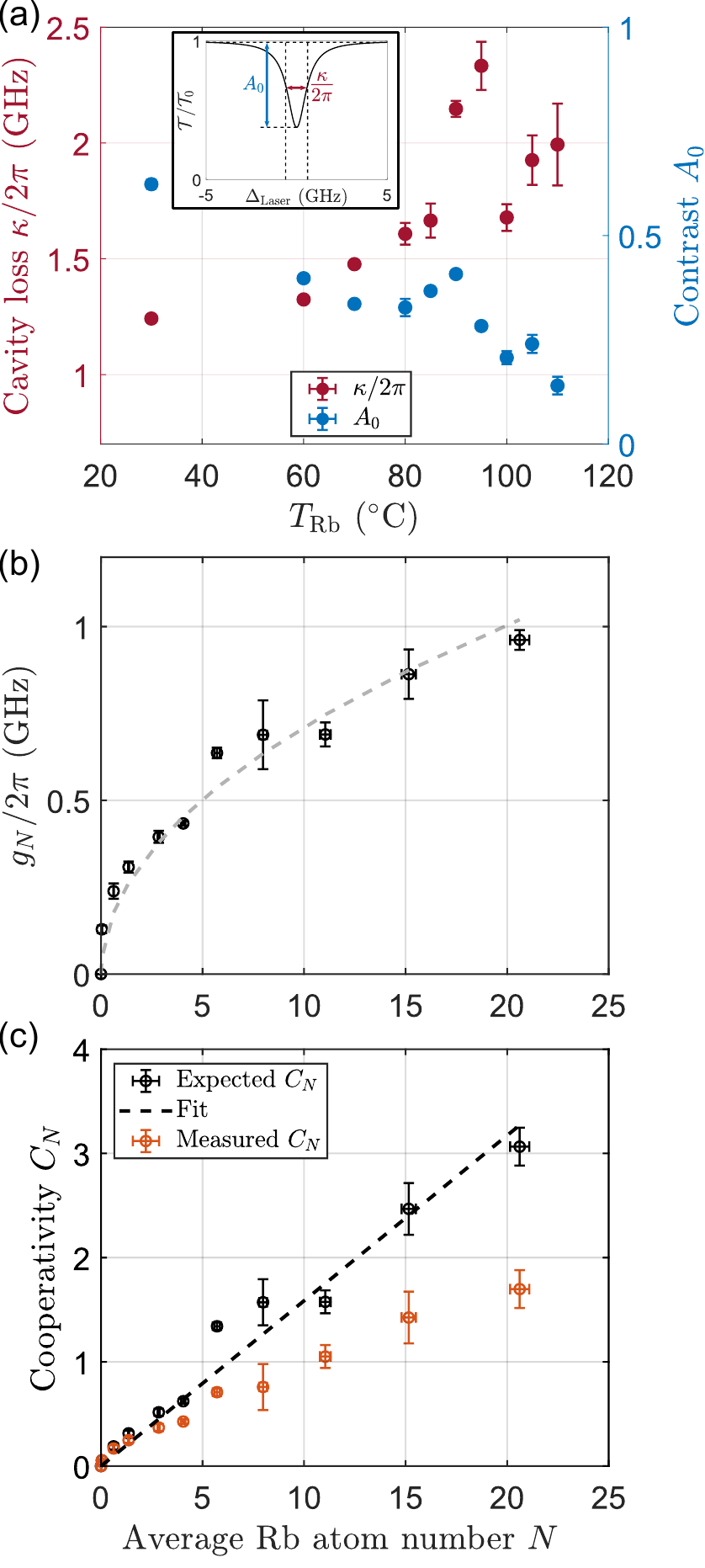}
  \caption{
    \textbf{(a)} The temperature dependence of cavity-photon loss rate $\kappa$ (in red) and the resonance contrast $A_0$ (in blue). Inset shows explicitly how $\kappa$ and $A_0$ are extracted from the bare-cavity resonance profile.
    \textbf{(b)} Measured collective coupling strength $g_N$. The gray dashed line shows a fit with $g_N = g_0\sqrt{N}$. This panel shares a common x-axis with panel (c).
    \textbf{(c)} Calculated cooperativity with $C_N=\frac{g_N^2}{\kappa\gamma}$. Black data correspond to $C_N$ with $\kappa$ at $T_{\mathrm{Rb}}=\SI{30}{\celsius}$, black dashed line is a linear fit to the black data with $C=NC_0$ and orange data show $C_N$ with $\kappa=\kappa(T_{\mathrm{Rb}})$.
    }
\label{fig:Splitting&Cooperativity}
\end{figure}

When the MRR is tuned a few GHz off resonance from the Rb transitions, it approaches the bare-cavity regime. We observe that the resonance linewidth broadens while the contrast decreases with increasing $T_{\mathrm{Rb}}$, as shown in Fig.~\ref{fig:Splitting&Cooperativity}(a). The cavity photon decay rate $\kappa$ and contrast $A_0$ are extracted by fitting a single Lorentzian function to the normalized bare-cavity transmission spectrum; the error bars in Fig.~\ref{fig:Splitting&Cooperativity}(a) represent the fitting uncertainties. The increase in Rb vapor density introduces additional loss, likely due to scattering from Rb condensation or Rb diffusion on the MRR surface. This loss is reversible, and the original resonance lineshape is recovered upon reducing the vapor density by lowering the reservoir temperature. Consequently, measurements are limited to $T_{\mathrm{Rb}} \leq \SI{110}{\celsius}$, as at higher temperatures the resonance feature becomes barely discernible.

Figure~\ref{fig:Splitting&Cooperativity}(b),(c) summarize the measured coupling strength $g_N$ and cooperativity $C_N$ as functions of the average atom number $N$ at zero cavity detuning. Each data point is obtained from repeated measurements, with vertical error bars indicating systematic uncertainties and horizontal error bars reflecting the uncertainty in $N$ due to temperature fluctuations of $\pm \SI{0.5}{\celsius}$.
Figure~\ref{fig:Splitting&Cooperativity}(b) is fitted with $g_N = \sqrt{N}g_0$, yielding $g_0/2\pi \approx \SI{225}{\mega\hertz}$, consistent with the theoretical value at the maximum evanescent field (Fig.~\ref{fig:g_r_plot} in \hyperref[AppendixB]{Appendix~B}). Figure~\ref{fig:Splitting&Cooperativity}(c) shows the expected cooperativity $C = g_N^2/(\kappa\gamma)$ (black), calculated using $\kappa/2\pi = \SI{1.2}{\giga\hertz}$ and $\gamma/2\pi \approx \SI{240}{\mega\hertz}$ at $T_{\mathrm{Rb}} = \SI{30}{\celsius}$, together with a linear fit (black dashed line).
In practice, $\kappa$ varies with $T_{\mathrm{Rb}}$ (Fig.~\ref{fig:Splitting&Cooperativity}(a)), and $\gamma = \langle v_{\mathrm{Rb}} \rangle / \delta_{1/e}$ scales with $\sqrt{T_{\mathrm{Rb}}}$. Using $\kappa(T_{\mathrm{Rb}})$ and $\gamma(T_{\mathrm{Rb}})$ yields the measured cooperativity shown in orange in Fig.~\ref{fig:Splitting&Cooperativity}(c), giving $C_N \approx 2$ for $\sim 20$ $^{85}\mathrm{Rb}$ atoms.

\begin{figure}[htbp]
  \centering
  \includegraphics[width=0.75\columnwidth]{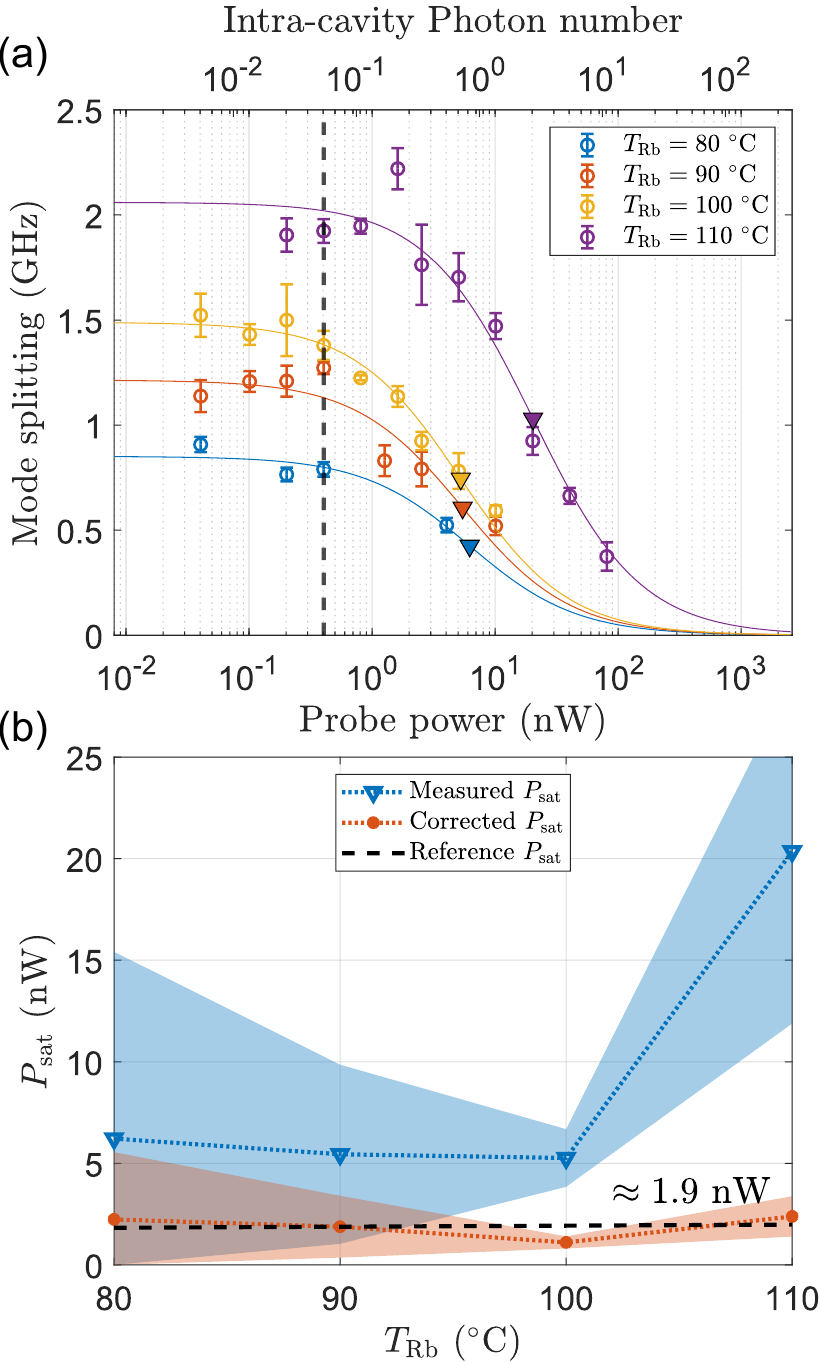}\par

  \caption{
    \textbf{(a)} Measured mode splitting depending on probe power inside the waveguide at different $T_{\mathrm{Rb}}$. Solid lines are fits using the saturation relation $S=S_0\left(\frac{1}{1+P/P_{\mathrm{sat}}}\right)$, where $S_0$ denotes the weak-probe splitting and $S$ is the power-dependent splitting~\cite{Zektzer:24}. Triangles denote the fitted saturation power for each $T_{\mathrm{Rb}}$. Black dashed line marks the probe power for measuring all cooperativity data.
    \textbf{(b)} Extracted saturation power from (a) (in blue) and corrected saturation power (in orange), with dotted lines as guidance to the eye and shaded region as the uncertainty of the fitting in (a). Black dashed line is the reference saturation power inferred with $\kappa(T_{\mathrm{Rb}}=\SI{30}{\celsius})$ and $\langle n \rangle_{\mathrm{sat}} \approx 0.74$.}
\label{fig:powerbroadening}
\end{figure}

The mode splitting at a given $T_{\mathrm{Rb}}$ depends on the probe power and follows the saturation behavior, as shown in Fig.~\ref{fig:powerbroadening}(a). This reflects a reduction in the number of atoms that respond linearly to the probe field.
Within our theory model, the collective coupling scales with the population difference $\langle \sigma^z \rangle$ (see Eq.~\eqref{eq:atomic_population} in \hyperref[AppendixC]{Appendix~C}). For weak excitation, $\langle \sigma^z \rangle \neq 0$, and atoms couple linearly to the cavity, producing mode splitting. As the probe power exceeds the saturation threshold, $\langle \sigma^z \rangle \to 0$, reducing the effective number of participating atoms; the system approaches the bare-cavity regime and the splitting vanishes.
The extracted saturation power at each $T_{\mathrm{Rb}}$ (triangles) is compared with the probe power of $\SI{0.45}{\nano\watt}$  (black dashed line) used for the cooperativity measurements, confirming operation well below saturation.

The fitted saturation power $P_{\mathrm{sat}}$ (blue) in Fig.~\ref{fig:powerbroadening}(b) increases from $\SI{6.5}{\nano\watt}$ at $T_{\mathrm{Rb}} = \SI{80}{\celsius}$ to above $\SI{20}{\nano\watt}$ at $\SI{110}{\celsius}$. To explain this trend, we estimate the intracavity photon number $\langle n \rangle_{\mathrm{sat}} = \gamma^2/(2g^2)$~\cite{HJKimble_1998} required to saturate an atom at the maximum evanescent field, with $g/2\pi = \SI{220}{\mega\hertz}$ and $\gamma = \langle v_{\mathrm{Rb}} \rangle / \delta_{1/e} \propto \sqrt{T_{\mathrm{Rb}}}$. Over $\SI{80}{\celsius} \le T_{\mathrm{Rb}} \le \SI{110}{\celsius}$, we find a weak increase of $\langle n \rangle_{\mathrm{sat}}$ from $0.71$ to $0.77$, allowing us to approximate $\langle n \rangle_{\mathrm{sat}} \approx 0.74$.

The corresponding input power follows from the steady-state solution of Eq.~\eqref{eq:intracavity_amplitude} in \hyperref[AppendixC]{Appendix~C},
\begin{equation}
    P_{\mathrm{sat}} = \langle n \rangle_{\mathrm{sat}} \hbar \omega \frac{\kappa^2}{4\kappa_{\mathrm{e}}},
\end{equation}
revealing a dependence on $\kappa^2/(4\kappa_{\mathrm{e}})$, which increases with $T_{\mathrm{Rb}}$ (Fig.~\ref{fig:Splitting&Cooperativity}(a)) and accounts for the observed rise in $P_{\mathrm{sat}}$.
To isolate this effect, we normalize each measured $P_{\mathrm{sat}}$ by $\kappa^2/(4\kappa_{\mathrm{e}})$ relative to its initial-state value at $T_{\mathrm{Rb}} = \SI{30}{\celsius}$. The normalized values (orange) collapse to $P_{\mathrm{sat}} \approx \SI{1.9}{\nano\watt}$, consistent with $\langle n \rangle_{\mathrm{sat}} \approx 0.74$.


\emph{Conclusion.—}
In this work, we demonstrate collective strong coupling on an integrated platform consisting of a high-$Q$ ($\approx 3\times10^5$) Si$_3$N$_4$ MRR coupled to thermal atoms in an anodically bonded vapor cell. Clear cavity-mode splitting confirms the strong-coupling condition $g_N^2/(\kappa\gamma) \ge 1$, with the extracted coupling strength and cooperativity following the expected scaling with atom number. We also develop a theoretical model that reproduces the experimental results (see \hyperref[AppendixC]{Appendix~C}). At $C_N \approx 2$, corresponding to $\sim 20$ $^{85}$Rb atoms, we infer a single-atom cooperativity of $C_0 \approx 0.1$.
These results mark progress toward the single-atom strong-coupling regime ($C_0 \ge 1$). Reaching this limit will require reduced mode volume, enhanced evanescent-field overlap, and higher $Q$, achievable with photonic crystal cavities~\cite{PhysRevLett.124.063602,Alaeian2020}, suspended microring geometries~\cite{10.1063/1.4707898}, and reduced surface scattering loss~\cite{Chen2026}.
Beyond the single MRR system demonstrated in this work, the compact footprint and inherent scalability of this platform make it particularly attractive for extension to coupled multi-cavity architectures and enable on-chip quantum networks with localized reconfiguration. Our work establishes thermal-atom–integrated photonic microcavities as a promising platform for cQED, with potential applications in on-chip quantum networks~\cite{Reiserer2022,Luo2023,Knaut2024}, quantum nonlinear optics~\cite{Faraon,Birnbaum2005,PhysRevLett.105.163601}, chip-scale quantum photonics~\cite{Aoki_2006, Srinivasan2007}, and precision spectroscopy~\cite{Stern2013, 10.1063/1.1787942}.


\emph{Funding.}
This project is funded by DARPA SAVaNT through contract number W911NF-24-1-0029 and Carl-Zeiss-Stiftung Center for Quantum Photonics. 


\emph{Acknowledgement.-}
We thank Markus Greul and Mathias Kaschel from IMS CHIPS for helping with chip fabrication, and Frank Schreiber from University of Stuttgart for making the Rb vapor cell. We also thank Gerhard Rempe for insightful discussions.

\emph{Disclosures.-}
The authors declare no conflicts of interest.

\emph{Data availability}
The data that support the findings of this article are openly available.~\cite{cheng_zenodo_data_2026}

\appendix
\section{Appendix A: Measurement setup and localized cavity tuning}
\label{AppendixA}

\begin{figure*}[htbp]
  \centering

  \includegraphics[width=0.8\textwidth]{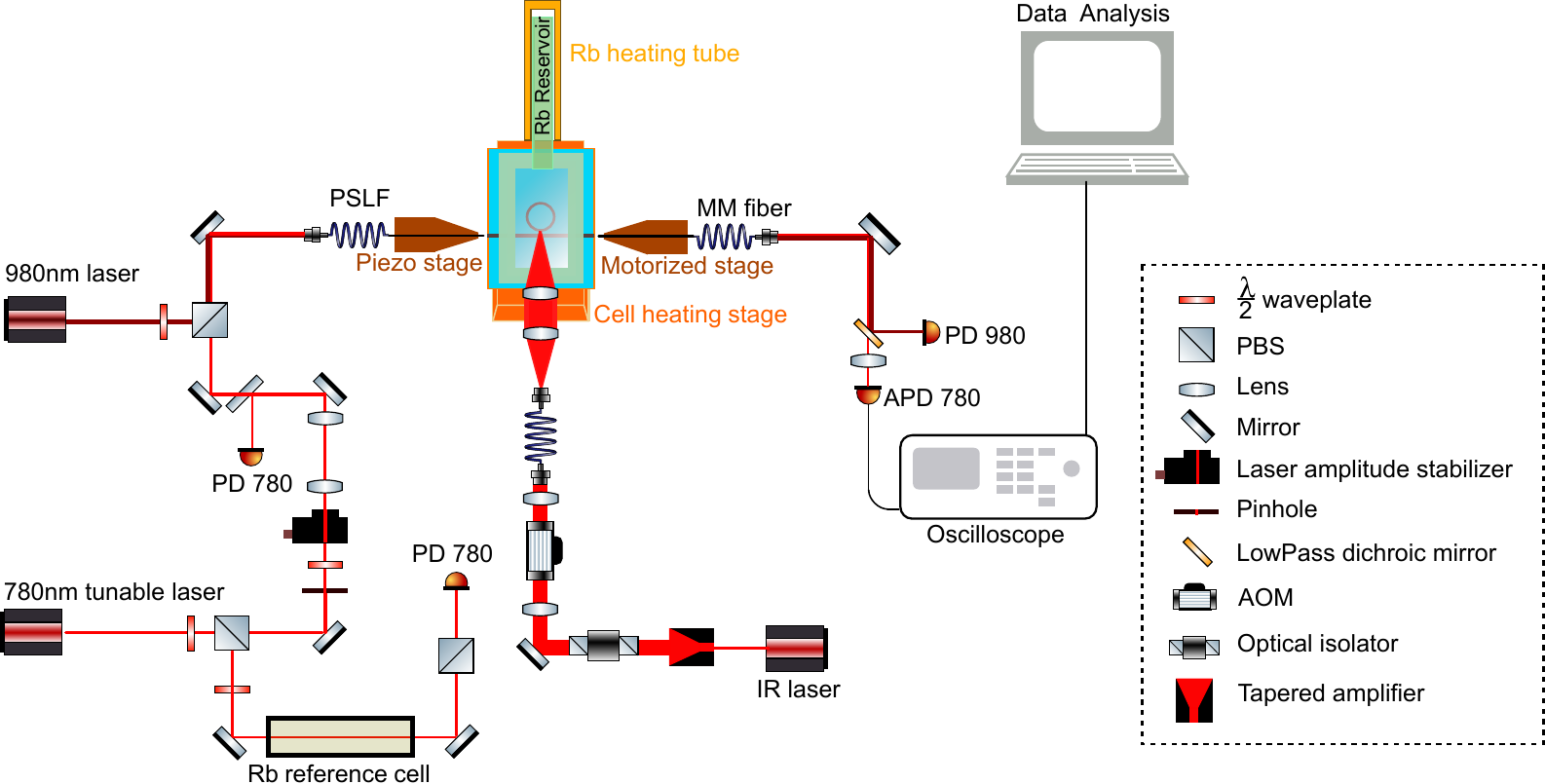}
  \caption{Schematic of the measurement setup. A $\SI{780}{\nano\meter}$ tunable laser sends in a weak probe laser while a $\SI{980}{\nano\meter}$ laser sends in a strong laser to stabilize the fiber coupling efficiency. An infrared laser locally tunes the resonance of the MRR.}
\label{fig:setup}
\end{figure*}

Figure~\ref{fig:setup} shows the schematic of the measurement setup. A tunable laser at $\SI{780}{\nano\meter}$ is split into two beams: one probes the transmission spectrum of a reference Rb cell, providing a frequency reference for the measurement, while the other, together with a tracking laser at $\SI{980}{\nano\meter}$, is coupled into a PSLF. The PSLF is mounted on a piezo-stabilized stage and couples light into the waveguide via edge coupling. A MM fiber collects the transmitted light, which is then separated by a dichroic mirror into the $\SI{780}{\nano\meter}$ signal and the $\SI{980}{\nano\meter}$ tracking signal, the latter serving as a feedback signal for the piezo-stabilized stage. The piezo-stabilized stage compensates the thermal drift and vibration, maintaining the PSLF coupling efficiency at an optimum level during the measurement. An infrared laser is amplified and tightly focused onto the MRR to locally tune the resonance via the thermo-optic effect, with its intensity modulated by an AOM. The resulting resonance tuning range is approximately $\SI{10}{\giga\hertz}$. For an AOM modulation period of $\SI{50}{\second}$, the MRR responds sufficiently rapidly to follow the IR-induced tuning with negligible hysteresis.

The local tuning rate of the MRR is chosen to be sufficiently slow ($\SI{0.4}{\giga\hertz/\second}$) compared with the probe-laser scan speed ($\SI{1}{\tera\hertz/\second}$), so that each $\SI{10}{\milli\second}$ scan can be treated as a time bin within which the cavity resonance shift is negligible. In this way, the cavity resonance may be regarded as effectively stationary during each individual probe scan. Further improvements in tuning speed could be achieved by integrating phase-change materials~\cite{Venkatachalam2011,Grim2019} or by exploiting the electro-optic response of the photonic material platform~\cite{Holzgrafe1586}, thereby enabling rapid local tuning and dynamic reconfiguration of MRR-based systems when needed.

\begin{figure*}[htbp]
  \centering
  \includegraphics[width=0.9\textwidth]{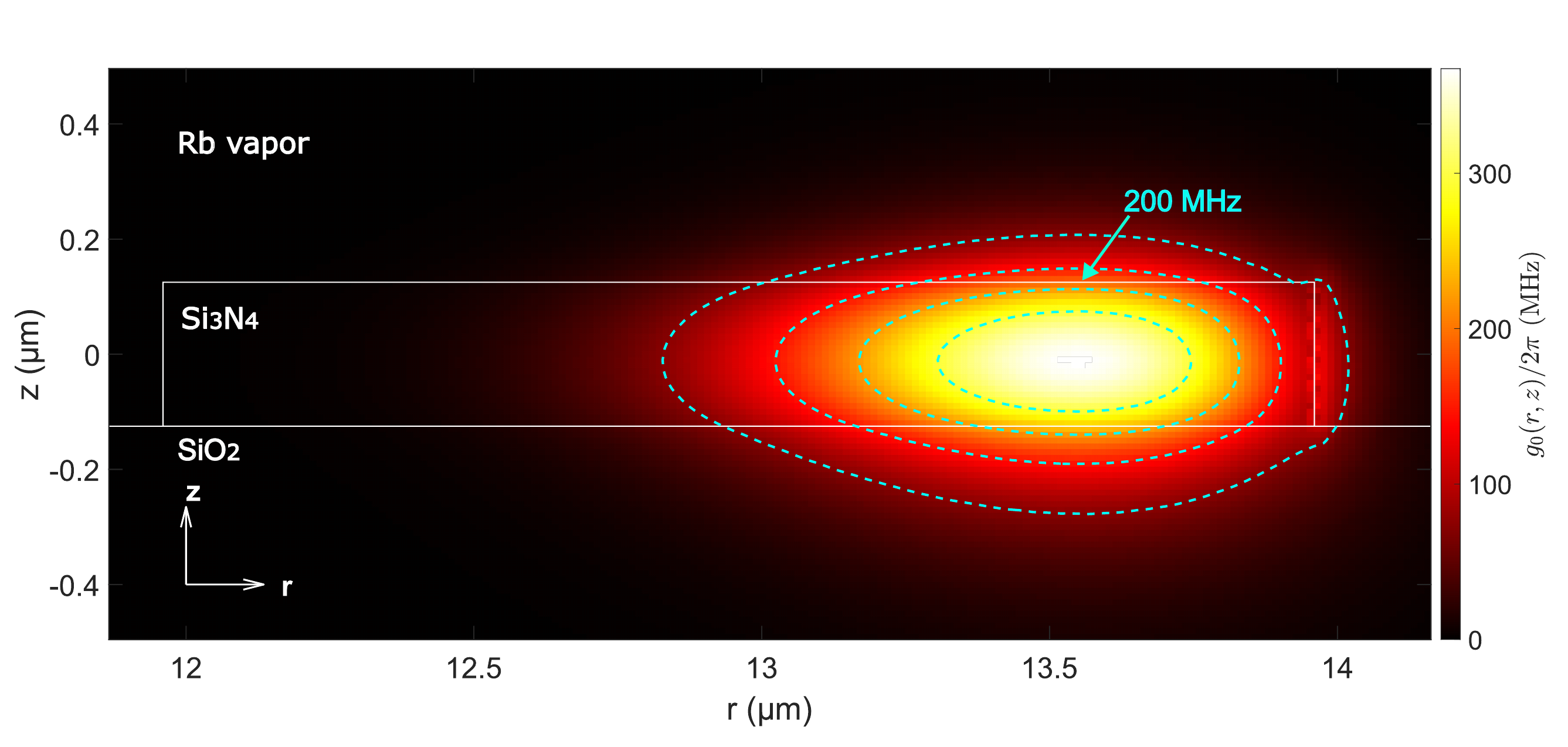}
  \caption{Calculated single-atom coupling strength $g_0(\mathbf{r})$. Dashed lines are contours of equal coupling strength. White solid lines indicate the waveguide cross-sectional boundaries and the substrate boundary, with the Si$_3$N$_4$ core and the $\mathrm{SiO_2}$ substrate beneath it, while the surrounding region above is filled with Rb vapor. The blue arrow indicates the location of the maximum single-atom coupling strength on the surface of the MRR.}
\label{fig:g_r_plot}
\end{figure*}

\begin{figure}[htbp]
  \centering
  \includegraphics[width=\columnwidth]{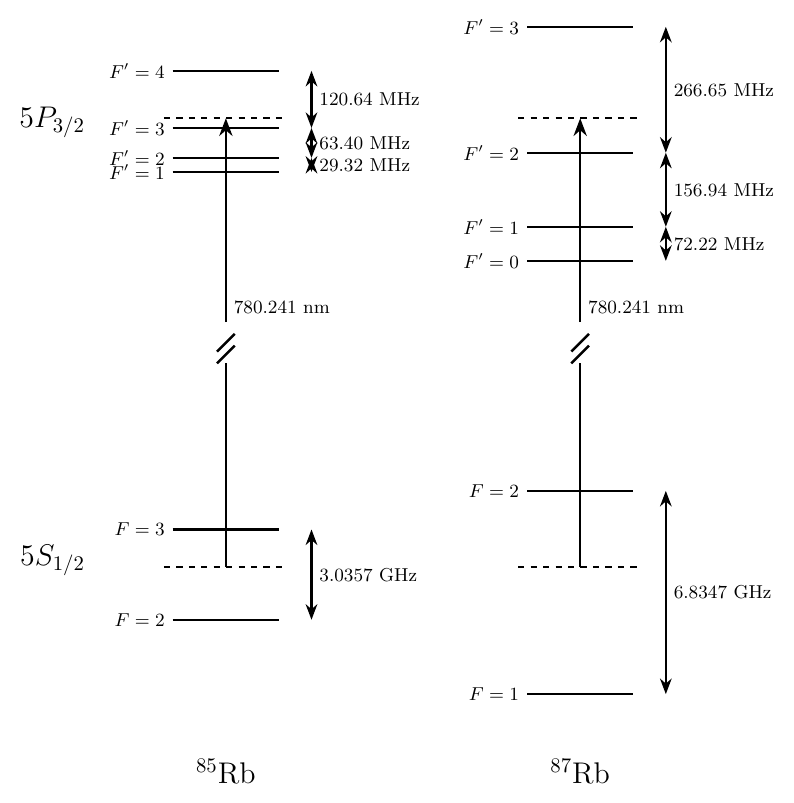}\par

  \caption{Hyperfine level structure of the $\mathrm{D_2}$ transition manifolds of $^{85}\mathrm{Rb}$ and $^{87}\mathrm{Rb}$~\cite{Steck_alkali}. The hyperfine excited-state splittings are smaller than the Doppler broadening and transient decoherence rate, resulting in the smearing and merging of avoided-crossings in Fig.~\ref{fig:2Dcolormap&slicing}(b).}
\label{fig:Rb_levelScheme}

\end{figure}

\section{Appendix B: Mode volume and average Rb atom number}
\label{AppendixB}
Thermal Rb vapor distributes atoms randomly around the MRR. Therefore, there is no predefined number of atoms inside the cavity mode. We estimate the average atom number from the evanescent-field mode volume inside the vapor, which is defined with the convention of using the mode energy density
\begin{equation}
  V_{\mathrm{evan.}}= \frac{\int_{\mathrm{vapor}} \varepsilon(\mathbf{r}') \left| E(\mathbf{r}') \right|^2 \, d\mathbf{r}'}{{\max\!\left(\varepsilon(\mathbf{r}) \left| E(\mathbf{r}) \right|^2\right)}_{\mathrm{vapor}}}.
\end{equation}

where $\varepsilon(\mathbf{r})$ and $E(\mathbf{r})$ are the position-dependent permittivity and electric-field amplitude.
Taking the maximum evanescent-field energy density in the vapor as the reference, we obtain $V_{\mathrm{evan.}}= \SI{2.54}{\micro\meter\cubed} $ . For a vapor density of $\SI{11.23}{\per\micro\meter\cubed}$ at $T_{\mathrm{Rb}}=\SI{110}{\celsius}$~\cite{PhysRev.27.578_Rb_density,PhysRevA.106.012801_Rb_density2}, the average number of Rb atoms is $\approx28.5$. Taking into account the natural abundance of $^{85}\mathrm{Rb}$ ($\approx\SI{72}{\percent}$), we obtain $N \approx 20$ $^{85}\mathrm{Rb}$ atoms coupled to the cavity evanescent field at zero cavity detuning. We estimate a single-atom coupling strength of $\bar{g}_0/2\pi\approx\SI{220}{\mega\hertz}$, consistent with the calculated coupling strength $g_0(\mathbf{r})$ at the location of maximum evanescent field in Fig.~\ref{fig:g_r_plot}. We calculate the electric-field distribution for a single intracavity photon and infer the position-dependent single-atom coupling strength using the following equations:

\begin{equation}
  g_0(\mathbf{r})=\frac{d}{\hbar}\sqrt{\frac{\hbar\omega}{2\varepsilon_0 V_{\mathrm{mode}}}}
\end{equation}
where $d$ is the averaged transition dipole moment of the $^{85}\mathrm{Rb}\ 5^2S_{1/2}\ F_g=2 \to 5^2P_{3/2}\ F_e=1,2,3$ hyperfine transitions, with linear polarized light, $\hbar\omega$ is the single-photon energy at a wavelength of $\SI{780}{\nano\meter}$, $\varepsilon_0$ is the vacuum permittivity, and $V_{\mathrm{mode}}$ is
\begin{equation}
  V_{\mathrm{mode}}= \frac{\int\varepsilon(\mathbf{r}')\left|E(\mathbf{r}') \right|^2 \, d\mathbf{r}'}{\max\!\left(\varepsilon(\mathbf{r}) \left| E(\mathbf{r}) \right|^2\right)}.
\end{equation}
where the reference energy density is chosen to be the maximum value of the total mode distribution.

\begin{figure}[h]
  \centering
  \includegraphics[width=0.75\columnwidth]{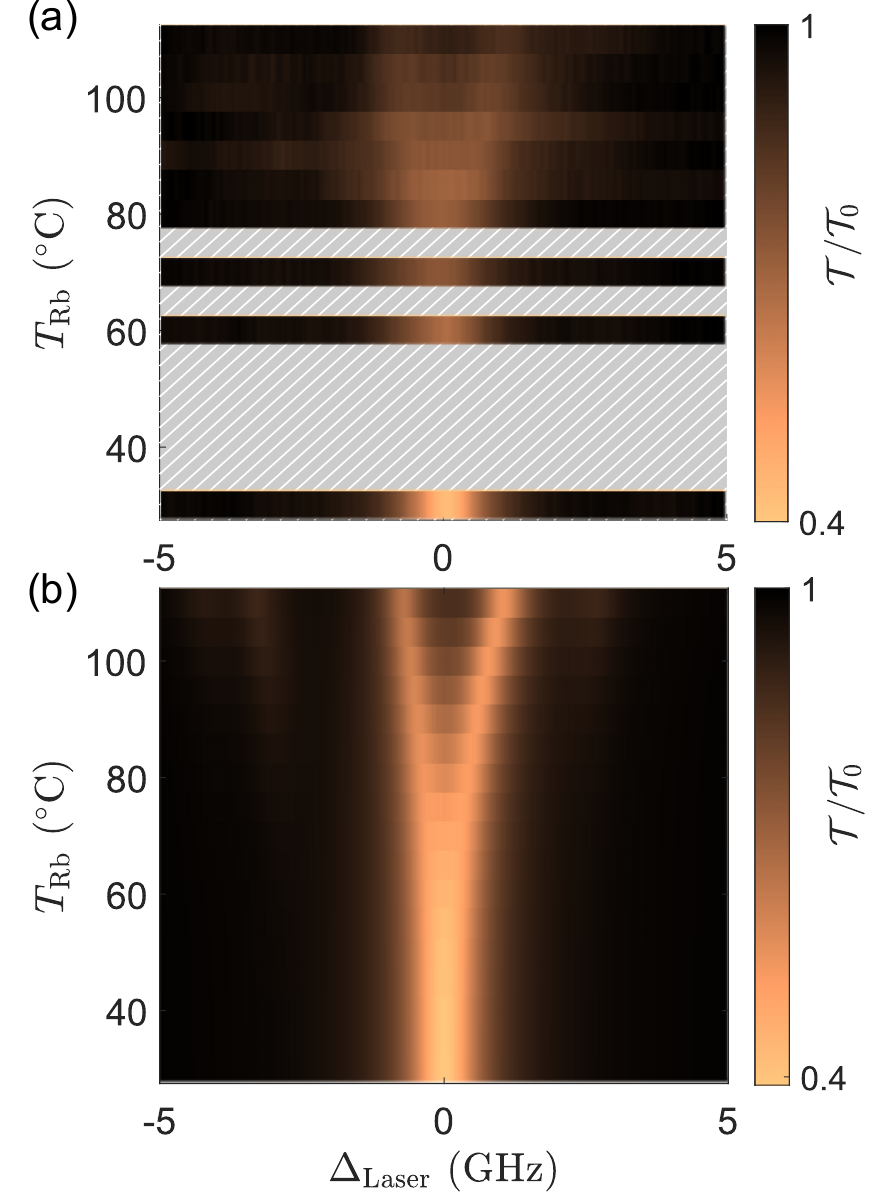}
  \caption{
    \textbf{(a)} Measured transmission spectra for different $T_{\mathrm{Rb}}$ at zero cavity detuning. Colorbars of (a) and (b) represent the normalized transmission $\mathcal{T}/\mathcal{T}_0$. For $T_{\mathrm{Rb}}\le\SI{80}{\celsius}$, hatched region denotes missing data.
    \textbf{(b)} Theoretically simulated transmission spectra at the same cavity--atom detuning for each $T_{\mathrm{Rb}}$.}
\label{fig:Theoretical_Simulation}
\end{figure}

\section{Appendix C: Theoretical model}
\label{AppendixC}
We model the atom–cavity interaction using the Tavis–Cummings Hamiltonian, and treat dissipation and coupling to external fields within the Gardiner–Collett input–output formalism~\cite{PhysRevA.31.3761}, in which the total Hamiltonian of our system is described by:
\begin{equation}
  H
  =H_{\mathrm{sys}}
  +
  H_{\mathrm{bath}}
  +
  H_{\mathrm{int}}.
\end{equation}
The closed cavity--atom system is described by the Tavis-Cummings Hamiltonian:
\begin{equation}
  H_{\mathrm{sys}}
  =
  \hbar \omega_{\mathrm{c}} a^\dagger a
  +
  \sum_{n=1}^{N}
  \hbar \omega_{\mathrm{a},n}\sigma_n^+\sigma_n^-
  +
  \sum_{n=1}^{N}
  \hbar g_n\left(a^\dagger \sigma_n^- + a \sigma_n^+\right),
\end{equation}
where $\omega_{\mathrm{c}}$ is the cavity frequency, $\omega_{\mathrm{a},n}$ is the atomic transition frequency of atom $n$, and $g_n$ is the corresponding atom--cavity coupling strength. Here, $a$ is the cavity-mode annihilation operator, and $\sigma_n^+$ and $\sigma_n^-$ are the raising and lowering operators of atom $n$.
The external baths consist of the bus-waveguide channel and the intrinsic cavity-loss channel:
\begin{equation}
  H_{\mathrm{bath}}
  =
  \hbar \int  \omega\, c^\dagger(\omega)c(\omega)\,d\omega
  +
  \hbar \int  \omega\, d^\dagger(\omega)d(\omega)\,d\omega\,,
\end{equation}
where $c(\omega)$ and $d(\omega)$ are annihilation operators at angular frequency $\omega$ for the bus-waveguide continuum mode and intrinsic-loss bath mode, respectively.
The interaction Hamiltonian between the closed system and the baths is
\begin{equation}
  \begin{aligned}
    H_{\mathrm{int}}
    &=
    i\hbar \int  \kappa_{\mathrm{e}}(\omega)
    \left[c^\dagger(\omega)a-a^\dagger c(\omega)\right] d\omega\\
    &\quad +
    i\hbar \int  \kappa_{\mathrm{i}}(\omega)
    \left[d^\dagger(\omega)a-a^\dagger d(\omega)\right]d\omega\,.
  \end{aligned}
\end{equation}
Here, $\kappa_{\mathrm{e}}(\omega)$ is the amplitude coupling between the cavity and the bus continuum, and $\kappa_{\mathrm{i}}(\omega)$ is the amplitude coupling between the cavity and the intrinsic-loss bath.
Under the Heisenberg picture, we derive the equations of motion from the Heisenberg--Langevin equations. With the rotating-wave approximation, Markovian approximation and mean-field approximation (valid in our strongly decoherent system), the equations in the laboratory frame reduce to
\begin{equation}
  \dot a
  =
  -\left(\frac{\kappa}{2}+i\omega_{\mathrm{c}}\right)a
  -i\sum_{n=1}^{N} g_n \sigma_n^-
  +
  \sqrt{\kappa_{\mathrm{e}}}\,a_{\mathrm{in}}
  +
  \sqrt{\kappa_{\mathrm{i}}}\,a_{0,\mathrm{in}},
\label{eq:intracavity_amplitude}
\end{equation}
\begin{equation}
  \kappa \equiv \kappa_{\mathrm{e}}+\kappa_{\mathrm{i}},
  \\
  \kappa_{\mathrm{e}}
  =
  2\pi
  \left|
  \kappa_{\mathrm{e}}(\omega_{\mathrm{c}})
  \right|^2
  \rho(\omega_{\mathrm{c}}),
\end{equation}
\begin{equation}
  \dot{\sigma}_n^- =
  -\left(\frac{\gamma_n}{2}+i\omega_{\mathrm{a},n}\right)\sigma_n^-
  +
  i g_n a \sigma_n^z,
\label{eq:atomic_population}
\end{equation}
\begin{equation}
  a_{\mathrm{out}} = a_{\mathrm{in}} - \sqrt{\kappa_{\mathrm{e}}}\,a.
\end{equation}

Here, $\kappa_{\mathrm{e}}$ is the observable cavity energy-decay rate into the bus waveguide, and $\kappa_{\mathrm{i}}$ is the intrinsic cavity-loss rate. $\rho(\omega_{\mathrm{c}})$ is the bath density of states. $a_{\mathrm{in}}$ is the input field operator oscillating at the probe frequency $\omega_{\mathrm{p}}$ in the bus waveguide, and $a_{0,\mathrm{in}}$ is the vacuum-noise input operator (which vanishes in expectation-value calculations). $\gamma_n$ is the atomic polarization decay rate of atom $n$, including spontaneous decay and transient effects. $\sigma_n^z$ is the Pauli-$z$ operator, whose expectation value gives the population difference $\langle \sigma^z \rangle=\rho_{\mathrm{ee}}-\rho_{\mathrm{gg}}$, where $\rho_{\mathrm{ee}}$ and $\rho_{\mathrm{gg}}$ are the excited- and ground-state populations, respectively. Under weak excitation, $\sigma_n^z\approx-1$. $a_{\mathrm{out}}$ is the output field operator in the bus waveguide. The probe laser is introduced as a coherent input in the waveguide channel within the input–output formalism, such that the input operator is replaced by its expectation value $a_{\mathrm{in}} \to \alpha_{\mathrm{in}} e^{-i\omega_{\mathrm{p}} t}$ in the Heisenberg–Langevin equations, $\omega_{\mathrm{p}}$ as the probe laser frequency. Here we select a single cavity mode by choosing the helicity set by the propagation direction of light in the waveguide.

By solving the steady-state expectation values of these operator equations in the rotating frame, we obtain the bus-waveguide transmission function:
\begin{equation}
  \begin{aligned}
\label{TransmissionFunction}
    T(\omega_{\mathrm{p}})
    =\left|\frac{\alpha_{\mathrm{out}}}{\alpha_{\mathrm{in}}}\right|^2
    =
    \left|
    1
    -
    \frac{\kappa_{\mathrm{e}}}
    {
      \frac{\kappa}{2}-i\Delta_{\mathrm{c}}
      +
      \displaystyle\sum_{n=1}^{N}
      \frac{g_n^2}{\frac{\gamma_n}{2}-i\Delta_{\mathrm{a},n}}
    }
    \right|^2.
  \end{aligned}
\end{equation}
with
\begin{equation}
  \Delta_c \equiv \omega_{\mathrm{p}}-\omega_{\mathrm{c}},
  \qquad
  \Delta_{a,n} \equiv \omega_{\mathrm{p}}-\omega_{\mathrm{a},n}.
\end{equation}

Notably, a similar transmission function can be derived from a classical analog model of two linearly coupled damped oscillators. In the weak-excitation limit, the Tavis--Cummings model becomes linear; therefore, a coherently driven quantum system evolves equivalently to a system of two classically coupled linear oscillators. As a result, transmission spectroscopy of this system does not require treatment beyond a linear dispersive model, as has been studied in the 1990s with atomic beams~\cite{CarmichaelNMSplitting1,CarmichaelNMSplitting2,RempeNMSplitting}.
We simulate the transmission spectrum with the derived transmission function (Eq.~\ref{TransmissionFunction}) using Monte Carlo sampling of atomic positions and thermal velocities to include Doppler shifts and transient decoherence, together with the position-dependent single-atom coupling strength $g_0(\mathbf{r})$ for each atom. We plot the simulated transmission spectrum at the same cavity--atom detuning as in Fig.~\ref{fig:2Dcolormap&slicing}(c) for each $T_{\mathrm{Rb}}$, and compare with the measured spectra, as shown in Fig.~\ref{fig:Theoretical_Simulation}. The simulations are in good agreement with the measurements and reveal the asymmetry of the mode splitting due to the off-resonance collective coupling from other hyperfine transitions.


\bibliographystyle{apsrev4-2}
\bibliography{references}

\end{document}